\documentclass[%
reprint,
superscriptaddress,
 amsmath,amssymb,
pra,
floatfix,
]{revtex4-2}

\usepackage{xcolor} 
\usepackage{graphicx}
\usepackage{dcolumn}
\usepackage{bm}

\usepackage[english]{babel}
\usepackage[autostyle=true]{csquotes}

\usepackage{hyperref}

\usepackage[mathlines]{lineno}

\begin{document}

\preprint{APS/123-QED}

\title{Superconducting Spiral Inductors for RF Reflectometry: Operation at Elevated Temperatures and Magnetic Fields}

\author{Euan Parry}%
\email{Email: euan.parry@strath.ac.uk}
\affiliation{
Department of Physics, SUPA, University of Strathclyde, Glasgow G4 0NG, United Kingdom}
\author{Murat Cubukcu}%
\affiliation{National Physical Laboratory, Hampton Road, Teddington TW11 0LW, United Kingdom}
\affiliation{London Centre for Nanotechnology, University College London, London, UK}
\author{Patrick Reuvekamp}%
\affiliation{National Physical Laboratory, Hampton Road, Teddington TW11 0LW, United Kingdom}
\author{Manoj Stanley}%
\affiliation{National Physical Laboratory, Hampton Road, Teddington TW11 0LW, United Kingdom}
\author{Jonathan D. Fletcher}%
\email{Email: jonathan.fletcher@npl.co.uk}
\affiliation{National Physical Laboratory, Hampton Road, Teddington TW11 0LW, United Kingdom}
\author{Alessandro Rossi}
\affiliation{
Department of Physics, SUPA, University of Strathclyde, Glasgow G4 0NG, United Kingdom}
\affiliation{National Physical Laboratory, Hampton Road, Teddington TW11 0LW, United Kingdom}

\date{\today}%

\begin{abstract}

Superconducting spiral inductors are emerging as key components for radio-frequency (RF) reflectometry, a widely used readout technique for semiconductor spin qubits. Future scalable quantum-computing architectures are expected to operate at elevated temperatures and magnetic fields, placing new demands on the performance and stability of superconducting circuit elements. Here, we present a systematic study of NbTiN spiral inductors under temperatures of several kelvin and magnetic fields approaching 1 T. By combining weakly coupled resonator measurements with independent two-port inductance extraction, we separate inductive and capacitive contributions to device behaviour and directly identify the origin of resonance shifts and quality factor degradation. Furthermore, we establish practical design metrics linking geometry, temperature sensitivity, and magnetic-field robustness. These results provide a general framework for benchmarking superconducting inductors and guiding the design of future RF-reflectometry circuits for practical quantum technologies.

\end{abstract}

\maketitle


\section{\label{sec:intro}Introduction}

\begin{figure}[t]
     \includegraphics[scale=1.1]{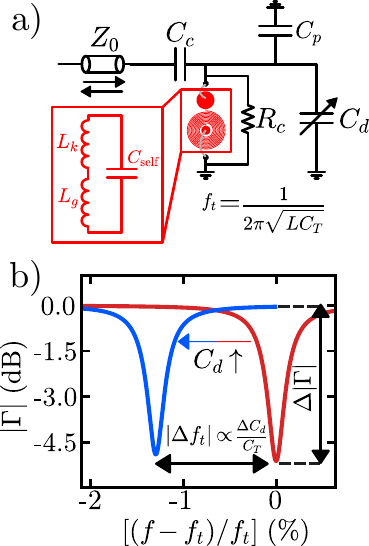}
\caption{\label{fig:1} Parallel configuration tank circuit for gate-based reflectometry. (a) Equivalent circuit showing the superconducting spiral and its sub-components (red) placed in parallel to a variable capacitor ($C_\textup{d}$) representing a quantum device. (b) Simulated reflection coefficient amplitude as a function of fractional resonant frequency shift. The traces illustrate a transition from low (red) to high (blue) device capacitance. The phase response is omitted for simplicity.}
\end{figure}
\begin{table*}[t]\label{tab:spiral_devices}
\centering
\begin{tabular*}{\textwidth}{@{\extracolsep{\fill}} c c c c c c c c c c c c c c @{}}
\hline
Device & $d_{\mathrm{out}}$ & $n$ & $w$ & $g$ & Length & $L_g^{\rm calc}$ & $L_k^{\text{calc}} (2~\text{K})$ & $L^{\text{calc}} (2~\text{K})$ & $L^{\rm meas}(2 K)$ & $\alpha=$ & $f_0 (2~\text{K})$ & $C_{\mathrm{self}}$ & $Q_i$ \\
label & $(\mu\mathrm{m})$ &  & $(\mu\mathrm{m})$ & $(\mu\mathrm{m})$ & $(\mathrm{mm})$ & $(\mathrm{nH})$ & $(\mathrm{nH})$ & $(\mathrm{nH})$ &  $(\mathrm{nH})$ & $L_k/(L_g+L_k)$ & $(\mathrm{GHz})$ & $(\mathrm{fF})$ & $(10^4)$ \\
\hline
$D_{1}$ & 500 & 15 & 6 & 6 & 14.8 & 70 & 4.3 & 74 & - & 0.058 & 2.61 & 50$\pm1$& 3.66 \\
$D_{2}$ & 430 & 21 & 4 & 4 & 17.0 & 103  & 7.4 & 110 & 113$\pm 2$ & 0.067 & 2.26 & 45$\pm1$& 3.59 \\
$D_{3}$ & 500 & 25 & 4 & 4 & 23.3 & 166 & 10.1 & 176 & - & 0.057 & 1.73 & 48$\pm1$ & 3.17 \\
$D_{4}$ & 450 & 12 & 6 & 6 & 11.3 & 47 & 3.3 & 50 & 50$\pm 2$ & 0.065 & 3.24 & 48$\pm1$ & 3.97 \\
$D_{5}$ & 330 & 25 & 3 & 3 & 13.9 & 87 & 8.1 & 95 & - & 0.085 & 2.93 & 31$\pm1$ & 4.20 \\
$D_{6}$ & 560 & 19 & 6 & 6 & 19.4 & 104 & 5.6 & 110  & 110$\pm 2$ & 0.051 & 2.11 & 52$\pm1$ & 3.44 \\
\hline
\end{tabular*}
\caption{Spiral device parameters and extracted zero-field resonator properties at $2~\mathrm{K}$. The geometric parameters, calculated geometric inductance $L_g$, and number of squares $N_{\square}$ are defined from the designed spiral geometries described in Sec.~\ref{sec:meth}. The kinetic inductance $L_k=N_{\square}L_{k,\square}$ was calculated using the sheet kinetic inductance $L_{k,\square}=1.74~\mathrm{pH/sq}$ extracted from the inductance measurements with set-up 2. The resonant frequency $f_0$ and internal quality factor $Q_i$ were extracted from the inductively coupled notch measurements with set-up 1.
}
\end{table*}
Radio-frequency (RF) reflectometry is widely used for high-bandwidth readout of quantum devices including spin qubits~\cite{schoelkopf1998rfset,ahmed2018rfcapacitive,vigneau2023rfreflectometry}.  Typically, a quantum device is embedded in a resonant tank circuit where the resonant frequency ($f_t$), bandwidth, and sensitivity are set by the total inductance, capacitance, and circuit losses. Surface-mount inductors are commonly used for the tank circuit because they provide convenient inductance values in the $10$--$1000~\mathrm{nH}$ range, placing typical readout signals in the $100~\mathrm{MHz}$ to few-GHz band~\cite{vigneau2023rfreflectometry}. However, such commercial inductors can be bulky, and can present significant parasitic capacitance and internal losses which limit readout sensitivity.\\\indent
An alternative that is gaining widespread adoption is superconducting spiral inductors~\cite{bugu2021sensitivity,denisov2023dispersive,vigneau2023rfreflectometry}. These provide a compact geometry in which the inductance, self-capacitance ($C_\textup{self}$), and coupling can be engineered lithographically. Their total inductance contains both a geometric contribution ($L_\textup{g}$), set by the spiral layout, and a kinetic contribution ($L_\textup{k}$), set by the inertia of the superconducting condensate. The use of superconducting materials suppresses dissipative loss compared with normal-metal inductors, while the planar spiral geometry allows large inductance to be achieved in a small footprint. Disordered superconductors such as NbN, NbTiN, WSi, and TiN provide large kinetic inductance, allowing a target inductance to be reached with a reduced physical footprint~\cite{beilvert2024cpw,annunziata2010nanoinductors,maleeva2018granular,swift2025superinductor}. 
\\\indent To better appreciate the advantages of spiral inductors, a typical tank circuit in parallel configuration is shown in Fig.~\ref{fig:1}(a). The inductor typically contributes a small $C_{\mathrm{self}}$, and has an effective resistance $R_c$ in parallel with the inductive branch to represent internal dissipative loss. Additional parasitic capacitance ($C_p$) and the quantum device capacitance ($C_d$) contribute to the total capacitance ($C_T$). In the case of gate-based reflectometry~\cite{colless}, one is interested in reading out changes in $C_d$, which are more readily detected for smaller values of $C_T$ and higher loaded quality factor ($Q_\textup{L}$), as given by the expression of the reflection coefficient shift~\cite{vigneau2023rfreflectometry}
\begin{equation}
|\Delta\Gamma| \propto Q_L\frac{|\Delta C_d|}{C_T}.
\end{equation}
Figure~\ref{fig:1}(b) illustrates this in terms of both reflection coefficient and resonant frequency shift. In fact, a change in device capacitance shifts the tank resonance by $\Delta f_t/f_t \simeq -\Delta C_d/(2C_T)$, producing a corresponding change in the complex reflection coefficient.
Hence, in comparison to surface mount inductors, superconducting spirals can provide smaller parasitic capacitances, including contributions from $C_p$ and $C_{\mathrm{self}}$ and increased fractional frequency shifts for a given $\Delta C_d$, which in tandem with a higher $Q_\textup{L}$ at cryogenic temperature lead to better readout sensitivities.

Recently, the viable operating environment of semiconductor spin qubits has expanded. While spin qubits have traditionally been operated at dilution-refrigerator base temperatures, there is increasing interest in operation at temperatures above $1~$K and in magnetic fields of several hundred millitesla~\cite{yang2020onekelvin, Hamonic2026,Camenzind2022}. This change is motivated by the substantially larger cooling power available at elevated temperatures, which can exceed that at the mixing chamber by several orders of magnitude. The resulting thermal budget opens a pathway towards the integration of cryogenic control electronics, multiplexers, amplifiers, and other support circuitry in close proximity to the quantum processor, thereby reducing wiring complexity and facilitating scalable quantum-computing architectures~\cite{Gonzalez-Zalba2021}. These operating conditions, however, place additional demands on superconducting circuit elements. Elevated temperature increases the quasiparticle population and modifies the kinetic inductance of the superconducting film, while magnetic fields can introduce vortices and associated microwave dissipation. Both mechanisms can alter the resonance frequency and quality factor of the resonator, potentially degrading readout sensitivity.\\\indent
In this work, we present a systematic study of NbTiN superconducting spiral inductors under elevated temperatures and magnetic fields deliberately chosen to reflect the operating conditions increasingly targeted for scalable spin-qubit architectures. We investigate the behaviour of several spiral designs at temperatures of several kelvin and magnetic fields approaching one tesla, where quasiparticle generation, kinetic-inductance variations, and vortex-related effects can significantly influence circuit performance. By combining weakly coupled resonator measurements with independent two-port inductance extraction, we separate inductive and capacitive contributions to the device response and directly identify the origin of temperature- and field-dependent resonator performance deterioration. Our approach enables quantitative extraction of kinetic-inductance effects, reconstruction of resonator behaviour from independently measured inductance changes, and the establishment of practical design metrics linking geometry, operating temperature, and magnetic-field robustness. 

\section{\label{sec:meth}Methods}

\begin{figure}[t]
     \includegraphics[scale=0.86]{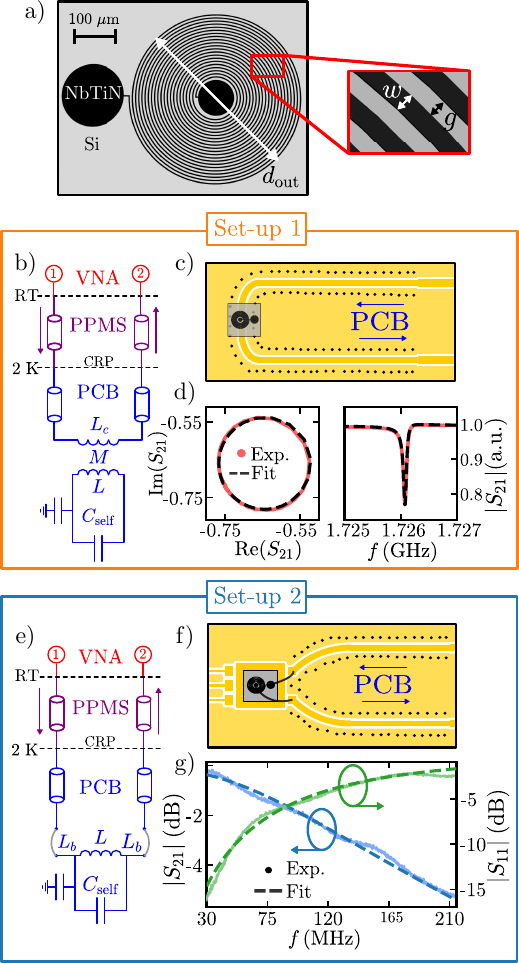}
\caption{\label{fig:2} Device geometry and microwave characterisation methods. (a) Schematic illustration of an individual spiral device and relevant geometric features. Inset: zoom-in showing track width and separation. (b) Equivalent circuit diagram for measurements carried out with set-up 1. The temperature gradient from room temperature (red) to cryogenic base temperature (blue) is shown. Dashed line labelled CRP indicates the reference plane of UOSM calibrations. (c) Schematic representation of a spiral chip placed above a CPW for the inductively coupled notch measurements enabled by set-up 1. (d) Experimental $S_{21}$ notch response as a function of frequency for device $D_3$ at $2~\mathrm{K}$ (red dot) and circle fits (black dashed). (e) Equivalent circuit diagram for measurements carried out with set-up 2. (f) Schematic representation of a spiral wire-bonded to a CPW for 2-port inductance extraction. (g) $S$-parameter experimental response (dot) and fit (dashed) as a function of frequency for device $D_2$ at $4~\mathrm{K}$.}
\end{figure}

\begin{figure*}[t]
\includegraphics{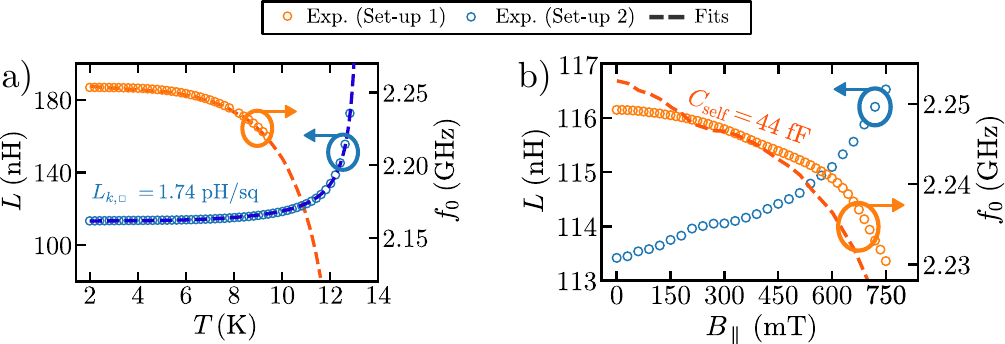}
\caption{\label{fig:3} (a) $L(T)$ extracted from the bonded two-port measurements of $D_2$ carried out in set-up 2 (blue circles) and $f_0(T)$ extracted from the inductively coupled notch measurements of a nominally identical $D_2$ device carried out in set-up 1 (orange circles). The blue dashed line is the fit discussed in Section~\ref{sec:inductance} and gives $L_g=106~\mathrm{nH}$, within $3\%$ of the designed value, and $L_{k,\square}=1.74~\mathrm{pH/sq}$. The orange dashed line shows the $f_0(T)$ calculated using Eq.~\ref{eq:resf} with the extracted inductance and a constant $C_{\mathrm{self}}=44~\mathrm{fF}$. (b) Extracted $L(B_\parallel)$ (blue circles) and $f_0(B_\parallel)$ (orange circles) for the same design. $B_\parallel$ is the magnetic field oriented nominally within the plane of the superconducting film. Using the same fixed $C_{\mathrm{self}}$, the orange dashed line represents $f_0(B_\parallel)$ calculated from Eq.~\ref{eq:resf}.}
\end{figure*}

\subsection{Spiral inductor devices}

The inductors consisted of $80~\mathrm{nm}$-thick NbTiN deposited and patterned on a high-resistivity silicon substrate and were fabricated by StarCryogenics [Fig.~\ref{fig:2}(a)]. The devices (labelled $D_{1}$--$D_6$) span a range of track widths $w$, turn numbers $n$, and outer (inner) diameter $d_{\text{out}}$ ($d_{\text{in}}$), as listed in Table~\ref{tab:spiral_devices}. These designs correspond to inductances in the tens to hundreds of nH relevant for use in RF reflectometry~\cite{vigneau2023rfreflectometry}.  

Spiral inductors behave as distributed resonators with inductance $L = L_g + L_k$ and $C_{\mathrm{self}}$ arising from proximity between concentric turns, giving a (self) resonance frequency
\begin{equation}\label{eq:resf}
f_0 = \frac{1}{2\pi\sqrt{LC_{\mathrm{self}}}}.
\end{equation}
Expressions used to estimate $L_g$ and $L_k$ are given in Appendix~\ref{supp:spiral_inductance}.

\subsection{Cryogenic set-up and calibration}\label{sec:cryo}

All measurements were performed in a Quantum Design Physical Property Measurement System (PPMS) with an adapted RF probe, enabling characterisation as a function of both temperature (from $2~$K) and applied magnetic field (up to $750~$mT). Microwave scattering parameters were measured using a vector network analyser. The RF probe was connectorised using coaxial RF sockets at the top of the probe and at the sample stage inside the PPMS.

Prior to measurement, an Unknown-Thru-Open-Short-Match (UOSM) calibration was performed using coaxial calibration standards at the lower end of the PPMS RF lines, following the unknown-thru two-port calibration method of Ferrero and Pisani~\cite{ferrero1992unknownthru}. This defined a calibrated reference plane close to the sample stage. The unknown-thru element was taken to be the coplanar waveguide PCB used in set-up~1, see Fig.~\ref{fig:2}(c). As such, the calibration reference plane was defined immediately at the sample environment in order to de-embed most of the spurious contributions due to the measurement apparatus.

Measurements were carried out at an input power of approximately $-50~\mathrm{dBm}$ at the device. At this drive level, two-level-system loss is expected to be largely saturated, which is appropriate for RF reflectometry and allowed the temperature- and field-dependent response of the superconducting film to be isolated from low-power dielectric loss mechanisms.

\subsection{Measurement techniques}

In reflectometry applications the spiral was wire-bonded to a single RF port at the end of a transmission line, corresponding to a one-port configuration, as in Fig.~\ref{fig:1}. However, for a thorough characterisation of the spirals in isolation, we adopted two independent but complementary measurement configurations that in combination allowed the contributions of inductance and self-capacitance to be separated without fitting ambiguity.

In set-up~1, the spiral chip was positioned above a coplanar waveguide (CPW) and coupled weakly to the transmission line [Fig.~\ref{fig:2}(b,c)]. This configuration forms a notch-type resonator, providing access to the intrinsic resonant response---including $f_0$ and $Q_i$---with only weak external loading~\cite{peruzzo2020geometric}. Complex $S_{21}$ responses were analysed using a circle-fitting routine~\cite{probst2015circlefit}; for device $D_3$ at $2~\mathrm{K}$ and zero field this yields $f_0 = 1.726~\mathrm{GHz}$ and $Q_i = (3.28 \pm 0.12)\times10^4$ [Fig.~\ref{fig:2}(d)].

In set-up~2, the spiral was wire-bonded between two RF ports [Fig.~\ref{fig:2}(e,f)], which allowed the inductance to be extracted independently from calibrated four-$S$-parameter measurements~\cite{pozar2011microwave}. The response was fitted over a low-frequency band where the structure can be treated as a lumped element, using an admittance model comprising a series $R$--$L$ branch with a small effective shunt capacitance to ground. The fitted resistance accounts for residual series loss in the spiral, bond wires, PCB traces, and measurement path, while the shunt capacitance captures the leading capacitive loading of the bonded fixture rather than the spiral self-capacitance. Corrections were then applied for bond-wire inductance contributions (Appendix~\ref{supp:bondwire}). For device $D_2$ at $4~\mathrm{K}$ this routine yields $L = 113 \pm 2~\mathrm{nH}$  as shown in Fig.~\ref{fig:2}(g) (see Appendix \ref{supp:bondwire} for details of sources of uncertainty). This technique was first validated by testing a commercial surface-mount inductor typically used in RF-reflectometry (see Appendix \ref{supp:coilcraft}).

By combining these two measurement routes, the resonance frequency could be obtained in two independent ways, i.e. from circle fits of set-up 1 measurements, as well as from calculations of Eq.~\ref{eq:resf} with inductance values extracted from  set-up 2 measurements and a constant $C_{\mathrm{self}}$ (assumed to be unaffected by $T$ and $B$ variations). This provided a direct test of whether observed frequency shifts were inductive or capacitive in origin. Further details of the fitting procedures are given in Appendices~\ref{supp:circle_fit} and~\ref{supp:sparam_L_fit}.

\begin{figure*}[t]
\hspace*{-0.1\columnwidth}
     \includegraphics{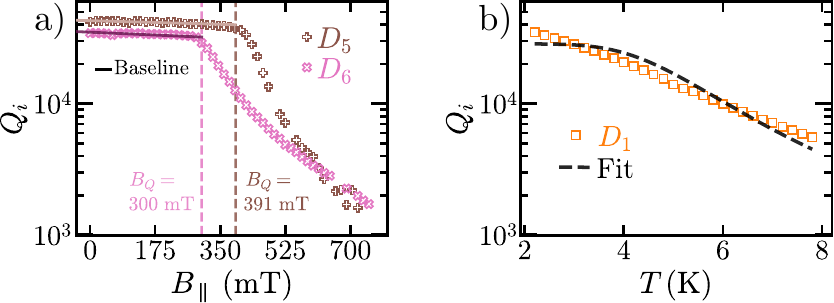}
\caption{\label{fig:4}Temperature- and field-dependent quality factor degradation. (a) $Q_i$ as a function of applied parallel magnetic field for devices $D_5$ and $D_6$ at $T = 2$~K.  The field values $B_Q=391~\mathrm{mT}$ for $D_5$ and $B_Q=300~\mathrm{mT}$ for $D_6$ for which a gradual degradation of more than 10\% manifests are highlighted by dashed vertical lines. (b) Temperature dependence of $Q_i$ for device $D_1$ in the absence of an applied field (squares). The dashed curve represents a fit to the quasiparticle-loss model of Eq.~\ref{eq:quas}.}
\end{figure*}
\section{\label{sec:inductance}Effects of Temperature and Field on inductance}

Figure~\ref{fig:3}(a) shows the temperature dependence of the inductance, $L$, and resonance frequency, $f_0$, of $D_2$, as extracted from our two separate measurement approaches. These measurements were performed on different physical devices with nominally identical design geometry. The inductance increase with temperature is consistent with the expected behaviour of a superconducting film, where an increased quasiparticle population reduces the superfluid density and enhances the kinetic inductance. The measured $L(T)$ was fitted using
\begin{equation}
L(T) = L_g + N_{\square}L_{k,\square}(T),
\end{equation}
where $N_{\square}$ is the number of squares of the spiral trace, defined from the designed geometry as the total trace length divided by the track width, and the temperature-dependent sheet kinetic inductance is given by \cite{tinkham2004superconductivity, annunziata2010nanoinductors}
\begin{equation}\label{eq:Lk}
L_{k,\square}(T) = \frac{hR_{\square}}{2\pi^2\Delta(T)}\coth\left(\frac{\Delta(T)}{2k_BT}\right),
\end{equation}
where $h$ is Planck's constant, $k_B$ is Boltzmann's constant, and $R_{\square}$ is the sheet resistance, defined for a uniform film of resistivity $\rho$ and thickness $t$ as $R_{\square}=\rho/t$, or equivalently through $R=R_{\square}l/w$ for a rectangular strip of length $l$ and width $w$. The gap $\Delta(T)$ is taken to follow the standard BCS temperature dependence of the superconducting gap~\cite{tinkham2004superconductivity}. The fit yields $L_g = 106~\mathrm{nH}$, within $3\%$ of the design value, together with $L_{k,\square}(0) = 1.74~\mathrm{pH/sq}$. Also extracted are the zero-temperature superconducting gap $\Delta_0 = 2.2~\mathrm{meV}$, and critical temperature $T_{c} = 13.4$ K, giving $\Delta_0/k_BT_c = 1.91$. Importantly, the extracted transition temperature is in agreement with the value we obtained from independent DC measurements, see Appendix~\ref{supp:dc_resistivity}.\\\indent In order to corroborate our working assumption that $C_{\mathrm{self}}$ remains approximately unaffected by temperature variations and largely defined by geometry alone, we used the values of $L$ and $f_\textup{0}$ extracted at $T=2~$K into Eq.~\ref{eq:resf}, which yielded $C_{\mathrm{self}} = 44~\mathrm{fF}$. We then kept $C_{\mathrm{self}}$ constant with temperature in Eq.~\ref{eq:resf}, and used $L(T)$ (blue dots) to calculate $f_\textup{0}(T)$, which is shown as the orange dashed trace in Fig.~\ref{fig:3}(a). The excellent agreement between the temperature dependencies of $f_\textup{0}$ evaluated in these two independent ways (orange dots vs dashed line) is a strong indicator that our assumption around $C_{\mathrm{self}}$ was sound.  More specifically, this agreement demonstrates that the dominant temperature-dependent shifts in the resonance frequency arise from inductive effects, while effects such as current redistribution as a function of temperature do not notably change $C_{\text{self}}$. Furthermore, given that measurements were performed on physically different devices, the agreement also indicates that device-to-device variation between nominally identical spirals is likely to be negligible.\\\indent
A similar analysis was applied to the magnetic-field dependence shown in Fig.~\ref{fig:3}(b), where $B_\parallel$ is defined as the magnetic field nominally oriented within the plane of the superconducting film (owing to unavoidable alignment uncertainties between experimental runs, a small residual out-of-plane component may be present). Again, the orange dashed line of Fig.~\ref{fig:3}(b) shows the expected $f_{0}(B_{\parallel})$ from Eq. \ref{eq:resf}, using the measured $L(B_{\parallel})$ (blue dots) in combination with a fixed $C_{\mathrm{self}} = 44~\text{fF}$  (the same value obtained from the temperature-dependent analysis).
Overall, the field effects on $f_0$ can be primarily attributed to inductive changes given the reasonable reported agreement between the behaviour observed  in the two measurement schemes. The observed deviation of the reconstructed $f_{0}(B_{\parallel})$ from the measured values (dashed line vs orange dots in Fig.~\ref{fig:3}(b)) is attributed to a small 
misalignment of the in-plane field, since the measurements of $f_0(B_{\parallel})$ and $L(B_{\parallel})$ were taken during separate experiments.\\\indent
These results show that the observed temperature- and field-dependent frequency shifts can be accounted for by changes in inductance alone. A single value of $C_{\mathrm{self}}$, obtained from the temperature-dependent data, is sufficient to reconstruct the resonance frequency shift in temperature and under applied magnetic field, within the uncertainty introduced by unintentional field misalignment between experimental runs. 

\section{\label{sec:loss}Microwave losses}
We next examine microwave loss under temperature and magnetic field. Specifically, we extracted $Q_i$ dependencies from inductively coupled notch measurements, i.e. using set-up 1. The field dependence for devices $D_5$ and $D_6$ at $2~\mathrm{K}$ is reported in Fig.~\ref{fig:4}(a), and the temperature dependence for device $D_1$ at zero field is in Fig.~\ref{fig:4}(b), see Appendix~\ref{supp:BT} for the full data-sets.\\\indent
    Figure~\ref{fig:4}(a) shows that $Q_i$ remains approximately constant for low applied in-plane fields before decreasing drastically above a device-dependent onset field. We define $B_Q$ as the field at which $Q_i$ decreases by $10\%$ from its low-field baseline. The exact criterion chosen is not critical to this analysis; it is a convenient threshold for the field scale that tracks the onset of measurable microwave loss due to vortex entry. The reduction in $Q_i$ above $B_Q$ is attributed to vortex-associated microwave loss, arising from residual perpendicular components of the applied field due to small angular misalignment. In practice, even small misalignment angles (here $\theta \sim 5^\circ$, as extracted from the fits in Sec.~\ref{sec:design}) produce a residual perpendicular component sufficient to nucleate vortices. $B_Q$ therefore provides a practical figure of merit for field robustness under realistic experimental conditions. We note that contributions from enhanced quasiparticle generation and vortex motion may also be present~\cite{song2009vortices,kwon2018fieldloss,zollitsch2019fieldreorientation,mcrae2020materialsLoss}.\\\indent
Figure~\ref{fig:4}(b) shows the temperature dependence of $Q_i$ in the absence of a magnetic field. As the temperature increases, $Q_i$ decreases, consistent with an increasing quasiparticle population and the associated increase in microwave surface resistance~\cite{mattis1958anomalous,tinkham2004superconductivity,gao2008thesis,zmuidzinas2012superconducting}. In the low temperature range of our experiments, the expected $Q_i$ saturation is not observed, suggesting that the low-temperature loss limit is well below $T=2$~K for these devices. The results can be described by the phenomenological quasiparticle-loss model \cite{mattis1958anomalous,zmuidzinas2012superconducting,devisser2011sparseqp}
\begin{equation}\label{eq:quas}
Q_i^{-1}(T) = Q_{\mathrm{other}}^{-1} + A\sqrt{T}\exp\left(-\frac{\Delta(T)}{k_BT}\right),
\end{equation}
where $Q_{\mathrm{other}}^{-1}$ represents approximately temperature-independent loss channels and the second term describes an activated quasiparticle contribution. The fit to the experimental data shown in Fig.~\ref{fig:4}(b) has essentially one free parameter ($A$) and yields $Q_{\mathrm{other}} = 2.8\times10^4$. One can see that the model captures the high-temperature degradation of $Q_i$ well, although the observed loss may additionally include contributions from dielectrics, vortices, radiation, and residual background~\cite{devisser2011sparseqp,mcrae2020materialsLoss}.\\\indent
The microwave-loss measurements show that NbTiN spiral inductors retain high quality factors above $10^3$ over a broad range of temperatures and magnetic fields, while revealing two distinct performance limits: quasiparticle-induced dissipation at elevated temperature and vortex-associated losses under applied magnetic field. These trends are consistent with previous superconducting spiral-resonator studies, which have shown that spiral geometry strongly influences microwave current distribution, mode structure, temperature response, and magnetic-field tolerance~\cite{zhuravel2012rfphotoresponse,ghamsari2012hts_spiral,medahinne2025spiral}. The relative importance of these mechanisms depends on both the operating environment and device geometry. This observation motivates the introduction of practical design metrics, enabling quantitative assessment of the trade-offs relevant to RF-reflectometry applications.

\section{\label{sec:design}Design implications for spiral inductors}

\begin{figure*}[t]
\hspace*{-0.1\columnwidth}
\includegraphics[scale=0.91]{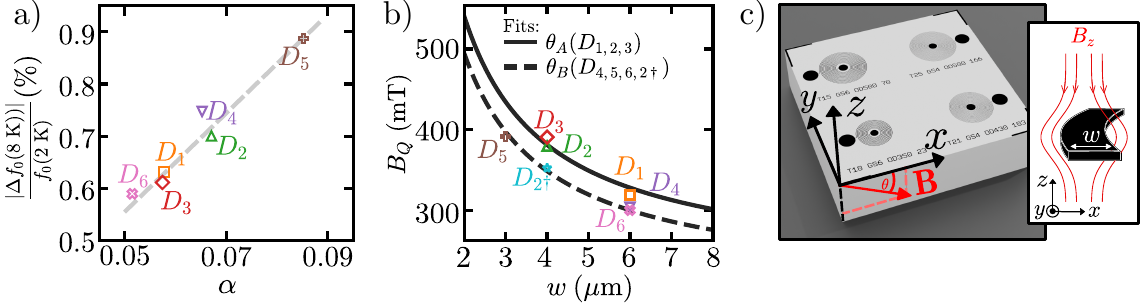}
\caption{\label{fig:5} Design metrics. (a) Fractional frequency shift between $2~\mathrm{K}$ and $8~\mathrm{K}$ as a function of kinetic-inductance fraction $\alpha=L_k/(L_g+L_k)$. Larger $\alpha$ gives greater temperature sensitivity, although all measured shifts remain below $1\%$. The grey dashed line is a guide to the eye. (b) Field-degradation threshold, $B_Q$, as a function of spiral track width, $w$. The dashed and solid curves are fits to the model represented by Eq.~\ref{eq:Bq}, which differ for the choice of effective alignment angles between different experimental runs. (c) Illustration of a chip used in this work containing 4 individual spiral designs. The magnetic field $\mathbf{B}$ may be unintentionally applied at an angle, $\theta$, from the $xy$-plane (angle significantly exaggerated for clarity) from one experimental run to the next. Inset: Illustration of the residual perpendicular component $B_z = |\mathbf{B}|\sin\theta$ which drives vortex entry at the strip edges; wider tracks are more susceptible to this perpendicular component, leading to the reduced $B_Q$ observed in panel (b).}
\end{figure*}

To link device performance to practical design choices, we now turn to study the kinetic-inductance fraction, $\alpha=L_k/(L_g+L_k)$, with which we quantify how strongly changes in $L_k$ affect the resonance frequency, and the field-degradation scale, $B_Q$, in terms of device track width, $w$.

Figure~\ref{fig:5}(a) shows the temperature-induced fractional frequency shift as a function of $\alpha$ (also listed in Table~\ref{tab:spiral_devices}). Here, $L_g$ is taken from the analytical design estimate, while $L_k = N_{\square}L_{k,\square}$ is calculated using the extracted sheet kinetic inductance and the designed number of squares. Devices with larger kinetic-inductance participation are expected to exhibit larger temperature-dependent resonance shifts~\cite{gao2006kineticfraction,gao2008thesis,zmuidzinas2012superconducting}, since to first order
\begin{equation}
\frac{\Delta f_0}{f_0}
\simeq
-\frac{\alpha}{2}
\frac{\Delta L_k}{L_k}.
\end{equation}
The results are consistent with this expectation. However, the measured shift between $2~\mathrm{K}$ and $8~\mathrm{K}$ remains below $1\%$ for all devices, indicating stable operation for RF reflectometry over this temperature range.

The choice of $\alpha$ is therefore a central design consideration. It can be increased geometrically by reducing the track width, which increases the number of squares and hence the kinetic-inductance contribution for a given footprint. It can also be increased through material choice, for example by using films with larger sheet kinetic inductance. Higher $L_{k,\square}$ can be obtained by reducing film thickness or increasing disorder, both of which increase the kinetic-inductance contribution per square~\cite{annunziata2010nanoinductors,bretzsullivan2022nbtin,frasca2023nbn,maleeva2018granular,swift2025superinductor}. This route is attractive for scalability because it allows a required inductance to be obtained with fewer turns and a smaller footprint. However, increasing $\alpha$ makes the resonator frequency more susceptible to changes in the superconducting state, so miniaturisation through high-$L_k$ designs must be balanced against operational stability under target qubit temperature and magnetic field operational constraints.
\\\indent Figure~\ref{fig:5}(b) shows the field-degradation scale, $B_Q$, as a function of $w$. The measured behaviour is consistent with vortex penetration physics (see also Appendix~\ref{supp:vortex}), resulting in devices with wider tracks exhibiting smaller $B_Q$. As noted in related studies of field-compatible superconducting resonators~\cite{roitman2024vortexloss}, microwave loss in nominally parallel field can be dominated by small residual perpendicular field components. In the present devices, the film thickness is much smaller than the track width, so a small out-of-plane misalignment produces a perpendicular component that can drive vortex entry at the strip edges, as illustrated in Fig.~\ref{fig:5}(c). The two fit traces shown in Fig.~\ref{fig:5}(b) use the same width-dependent fitting parameters but different magnetic field alignment angles to account for the fact that devices were measured in separate experimental runs. 
The fitted curves differ by an effective alignment change of approximately $\theta_A-\theta_B \approx 0.5^\circ$ around a nominal misalignment of $\simeq 5^\circ$, illustrating the sensitivity to small changes in sample alignment. 
The measured decrease of $B_\textup{Q}$ with increasing track width demonstrates that the magnetic-field robustness of superconducting spiral inductors can be deliberately engineered through geometry. Although NbTiN possesses a high intrinsic critical field, the practical magnetic-field tolerance of a spiral resonator is determined not only by material properties but also by design choices \cite{roitman2024vortexloss}.
\\\indent In brief, our work highlights a design trade-off between footprint, temperature stability, and magnetic-field resilience. Narrower tracks increase $N_{\square}$ for a given footprint, increasing $L_k$ and enabling greater miniaturisation (see Appendix \ref{supp:min_dout}). They are also expected to improve field resilience through the approximate $B_Q \propto 1/w$ scaling. Materials or film geometries with larger $L_{k,\square}$ provide an additional route to miniaturisation without relying only on increased spiral length and total footprint. However, increasing the kinetic-inductance contribution also increases $\alpha$, making $f_0$ more sensitive to temperature and field through changes in the superconducting state. The optimal geometry and film therefore depend on whether the application prioritises compactness, frequency stability, low parasitic capacitance, or operation in magnetic field.

\section{\label{sec:dis}Conclusion and Outlook}

We have developed a measurement framework for superconducting spiral inductors that separates inductive and capacitive contributions to their microwave response under elevated temperature and magnetic field. By combining weakly coupled resonator measurements with independent two-port inductance extraction, we show that the measured temperature- and field-dependent resonance shifts are predominantly inductive in origin and can be reconstructed from independently measured changes in $L$. This confirms the role of kinetic inductance in setting device behaviour even when the total inductance is dominated by its geometric contribution. More broadly, the work complements previous studies of superconducting microwave resonators by focusing on the inductor itself as the central circuit element, providing a route to benchmark inductance, self-capacitance, microwave loss, temperature sensitivity, and field robustness on equal footing.\\\indent
In the context of RF reflectometry, these results show that lithographic superconducting spiral inductors can provide large inductance, low self-capacitance, and internal quality factors well above the values required for typical externally coupled readout circuits. This makes them attractive for compact spin-qubit readout circuits operating at elevated temperature and magnetic field, where surface-mount inductors can add footprint, parasitic capacitance, and loss. Compared with more aggressive high-kinetic-inductance approaches such as granular films or superinductors~\cite{maleeva2018granular,swift2025superinductor}, the NbTiN spirals studied here occupy an intermediate design space: they provide compact, low-loss inductors with modest kinetic-inductance participation and correspondingly small frequency shifts over the measured temperature range. The framework presented here can be used to compare future inductor technologies and to choose the appropriate balance between miniaturisation, frequency stability, and magnetic-field resilience for scalable reflectometry readout architectures.

\begin{acknowledgments}
We wish to thank W. Wong for useful discussions. AR acknowledges support from the UKRI Future Leaders Fellowship Scheme (Grant agreement: UKRI1071). JDF, AR and MC acknowledge the support of the UK government department for Business, Energy and Industrial Strategy through the UK National Quantum Technologies Programme.
\end{acknowledgments}

\appendix

\section{\label{supp:spiral_inductance}Spiral inductance estimates}

The total spiral inductance values listed in Table \ref{tab:spiral_devices} are estimated as the sum of geometric and kinetic contributions. The geometric contribution $L_g$ is calculated from the designed spiral geometry using the modified Wheeler expression~\cite{mohan1999planarspiral,wheeler1928inductance},
\begin{equation}\label{eq:Lg_supp}
L_g =
\frac{\mu_0 n^2 d_{\mathrm{avg}}}{2}
\left[
\ln\left(\frac{2.46}{\rho}\right)
+
0.2\rho^2
\right],
\end{equation}
where $n$ is the number of turns,
\begin{equation}
    d_{\mathrm{avg}}=\frac{d_{\mathrm{out}}+d_{\mathrm{in}}}{2},
    \qquad
    \rho=\frac{d_{\mathrm{out}}-d_{\mathrm{in}}}{d_{\mathrm{out}}+d_{\mathrm{in}}}.
\end{equation}
Here $d_{\mathrm{out}}$ and $d_{\mathrm{in}}$ are the outer and inner spiral diameters, and $\rho$ is the fill factor.

The kinetic contribution is estimated from the sheet kinetic inductance of the NbTiN film $L_k = L_{k,\square} N_{\square}$ where $N_{\square}=l/w$ and $L_{k,\square}$ is the sheet kinetic inductance, $l$ is the total spiral trace length, and $w$ is the track width. The kinetic-inductance fraction is then defined as
\begin{equation}
    \alpha = \frac{L_k}{L_g+L_k}.
\end{equation}
These estimates are used in Table~\ref{tab:spiral_devices} to compare the designed geometric inductance, kinetic-inductance contribution, and total expected inductance of each spiral.

\section{\label{supp:circle_fit}Circle fitting routine}

The resonator frequency and quality factors were obtained from complex microwave transmission measurements using set-up 1 (see Fig. \ref{fig:2}(b--d)). Since the spiral devices were designed for integration into external RF reflectometry circuits, no on-chip feedline was included. For characterisation, the spiral chip was instead placed on top of a coplanar waveguide, allowing the spiral mode to couple inductively to the microwave line through the substrate. This configuration is shown schematically in Fig.~\ref{fig:2}(c).

Measurements were performed in the PPMS using set-up~1. The microwave response was calibrated to the reference plane labelled CRP in Fig.~\ref{fig:2}(b), reducing the influence of standing waves and background transmission features before this point. The resonance appears as a notch in the complex $S_{21}$ response, as shown for device $D_3$ in Fig.~\ref{fig:2}(d).

The measured resonance was fitted using the standard complex notch-resonator model~\cite{khalil2012asymmetric,gao2008thesis,probst2015circlefit}
\begin{equation}
S_{21}(f)
=
A e^{-2\pi i f\tau}
\left[
1-
\frac{Q_l e^{i\phi}}
{|Q_c|\left(1+2iQ_l x\right)}
\right],
\end{equation}
where $A=a e^{i\alpha}$ describes the complex background transmission, $\tau$ is the residual electrical delay, $\phi$ accounts for impedance mismatch and circle asymmetry, and
\begin{equation}
    x=\frac{f-f_0}{f_0}
\end{equation}
is the normalised detuning. The loaded, internal, and coupling quality factors are related by
\begin{equation}
    \frac{1}{Q_l}=\frac{1}{Q_i}+\frac{1}{Q_c}.
\end{equation}
Here $Q_i$ describes dissipation internal to the resonator, while $Q_c$ describes the coupling of energy from the resonator into the external microwave circuit.

From the circle fit, $f_0$, $Q_l$, $Q_c$, and the diameter-corrected internal quality factor $Q_i$ were extracted at each temperature or magnetic-field point. Parameter uncertainties were obtained from the local covariance matrix of the nonlinear least-squares fit, following the circle-fit approach of Probst \textit{et al.}~\cite{probst2015circlefit}. Specifically, the covariance matrix was estimated from the Jacobian $J$ of the complex residuals as
\begin{equation}
    \mathbf{C}_{\mathrm{fit}}
    =
    s^2
    \left(J^{T}J\right)^{-1},
\end{equation}
where $s^2$ is the residual variance. The quoted fit errors correspond to the square roots of the diagonal elements of $\mathbf{C}_{\mathrm{fit}}$, with derived quantities such as $Q_i$ propagated from the fitted $Q_l$ and $Q_c$ values. See Appendix~\ref{supp:BT} for full magnetic-field and temperature dependencies of these parameters.

\section{\label{supp:sparam_L_fit}Inductance extraction from admittance-matrix fitting}

The inductance was extracted by fitting the measured complex S-parameters to a lumped two-port admittance model using measurements taken in set-up 2 (see Fig. \ref{fig:2}(e--g)). Each measurement point consisted of four measured traces, $S_{11}$, $S_{21}$, $S_{12}$, and $S_{22}$, which are interpolated onto a common frequency grid. The measured S-matrix is described using a low-frequency lumped model consisting of a series impedance
\begin{equation}
Z_s(\omega) = R + i\omega L
\end{equation}
between the two ports, together with an effective shunt capacitance $C_g/2$ from each port to ground.

Here, $C_g$ represents the residual port-to-ground capacitance of the bonded two-port measurement configuration, including contributions from the PCB pads, short traces, wire-bond launch geometry, and any local capacitance from the bonded spiral terminals to the surrounding ground environment. It is therefore a fixture-dependent nuisance parameter used to capture the leading capacitive admittance visible in the low-frequency four-S-parameter measurement. It should not be interpreted as the spiral self-capacitance $C_{\mathrm{self}}$ used elsewhere in the paper. The latter is the effective distributed capacitance associated with the spiral resonant mode and is extracted independently from the measured resonance frequency using Eq.~\ref{eq:resf}. In contrast, $C_g$ only accounts for the low-frequency shunt loading present in the bonded inductance-extraction geometry.

The corresponding admittance matrix is
\begin{equation}
\mathbf{Y} =
\begin{pmatrix}
Y_s + Y_g/2 & -Y_s \\
-Y_s & Y_s + Y_g/2
\end{pmatrix},
\label{eq:Y_matrix_model}
\end{equation}
where
\begin{equation}
Y_s = \frac{1}{R+i\omega L},
\qquad
Y_g = i\omega C_g .
\end{equation}
Here $Y_s$ describes the series current path through the spiral, while $Y_g/2$ describes the effective shunt admittance from each port to ground in the bonded measurement fixture.

For a reference impedance $Z_0=50~\Omega$, the model S-parameters are obtained from
\begin{equation}
    \mathbf{S}
    =
    \left(\mathbf{I}-Z_0\mathbf{Y}\right)
    \left(\mathbf{I}+Z_0\mathbf{Y}\right)^{-1}.
    \label{eq:Y_to_S}
\end{equation}
Writing the elements of $\mathbf{Y}$ as $Y_{ij}$, this gives
\begin{align}
    A &= 1 + Z_0Y_{11}, &
    B &= Z_0Y_{12}, \\
    C &= Z_0Y_{21}, &
    D &= 1 + Z_0Y_{22},
\end{align}
with determinant
\begin{equation}
    \Delta = AD - BC .
\end{equation}
The fitted S-parameters are then
\begin{align}
    S_{11} &= \frac{(1-Z_0Y_{11})D + Z_0Y_{12}C}{\Delta}, \\
    S_{12} &= \frac{-(1-Z_0Y_{11})B - Z_0Y_{12}A}{\Delta}, \\
    S_{21} &= \frac{-Z_0Y_{21}D - (1-Z_0Y_{22})C}{\Delta}, \\
    S_{22} &= \frac{-Z_0Y_{21}B + (1-Z_0Y_{22})A}{\Delta}.
\end{align}
The parameters $R$, $L$, and $C_g$ were obtained by directly fitting these complex model S-parameters to the measured $S_{11}$, $S_{21}$, $S_{12}$, and $S_{22}$ over a low-frequency band (up to $200~$MHz, far below the typical resonant frequency $\sim 2~$GHz of these devices.) This procedure is the low-frequency lumped-element limit of the equivalent-circuit S-parameter methods used for inductor characterisation~\cite{naishadham2001smdinductors,zlebic2014indcalc}.

\section{\label{supp:bondwire}Bond-wire inductance and measurement uncertainty}
The aluminium bond-wire contribution was estimated using the standard straight-wire self-inductance approximation~\cite{terman1943radio}:
\begin{equation}
L_{\mathrm{b}}
\simeq
\frac{\mu_0 l}{2\pi}
\left[
\ln\left(\frac{2l}{r}\right)-1
\right],
\label{eq:bondwire_inductance}
\end{equation}
where $l$ is the bond-wire length, $r$ is the wire radius, and $\mu_0$ is the permeability of free space. For $50~\mu\mathrm{m}$-diameter aluminium bond wires, corresponding to $r=25~\mu\mathrm{m}$, Eq.~\ref{eq:bondwire_inductance} gives approximately $L_{\mathrm{b}}\simeq 1~\mathrm{nH/mm}$. We assigned a $\pm 20\%$ uncertainty to this correction in the extracted inductor values. Thus, for a typical total bond-wire length of $4~\mathrm{mm}$, a correction of $4 \pm 1~\mathrm{nH}$ was used. This, in combination with uncertainty arising from fitting due to imperfections with the de-embedding procedure, gives rise to a typical uncertainty of $\pm 2$--$3~$nH for these inductance measurements.

\section{\label{supp:coilcraft}Inductance extraction validation}

\begin{figure}[h]
\centering
\includegraphics[width=1\columnwidth]{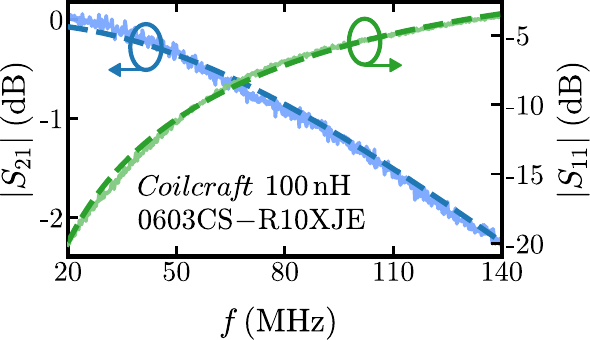}
\caption{\label{fig:supp_L}
Amplitude of S-parameter response as a function of frequency (solid lines) for a commercial surface mount inductor (Coilcraft $100~\mathrm{nH}$) acquired with set-up 2. Low-frequency admittance-matrix fits (dashed lines) enable the extraction of device inductance and comparison with commercial device datasheet for technique's validation.}
\end{figure}

The fitting procedure described in Appendix~\ref{supp:sparam_L_fit} was used to extract the inductance values shown in Fig.~\ref{fig:2}(g) and the temperature-dependent $L$ data in Fig.~\ref{fig:3}. As an independent validation of the extraction routine, the same measurement and fitting procedure was applied to a commercial Coilcraft $100~\mathrm{nH}$ inductor with a manufacturer-specified $5\%$ tolerance, as shown in Fig.~\ref{fig:supp_L}. The component was measured in set-up~2 using the PCB shown in Fig.~\ref{fig:2}(f), mounted on the dashed solder-pad footprint and wire-bonded to reproduce the spiral-device measurement configuration as closely as possible.

The fitted inductance of the Coilcraft component was $99\pm2~\mathrm{nH}$, consistent with its nominal $100~\mathrm{nH}$ value and within the datasheet tolerance. This agreement provides an independent check that the low-frequency admittance-matrix fitting procedure accurately extracts inductance values in the range relevant to the superconducting spiral devices.

\section{\label{supp:dc_resistivity}DC resistance measurement}

\begin{figure}[h]
    \centering
    \includegraphics[width=0.7\columnwidth]{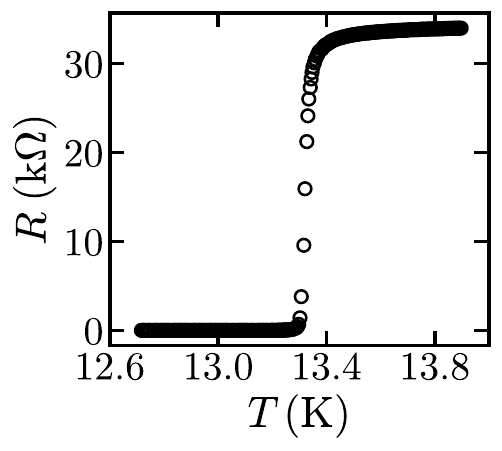}
    \hspace*{0.1\columnwidth}
    \caption{\label{fig:supp_dc_resistance}
    Two-point DC resistance measurement as a function of temperature. This is used to estimate the normal-state sheet resistance and superconducting transition temperature of the NbTiN spiral device.}
\end{figure}


A two-point source-measure-unit measurement was performed on a representative spiral device to estimate the normal-state sheet resistance of the NbTiN film. The measured device had a total line length of $11310~\mu\mathrm{m}$ and linewidth $6~\mu\mathrm{m}$, corresponding to $N_{\square} = 11310/6 \simeq 1885$.

The measured resistance as a function of temperature is shown in Fig.~\ref{fig:supp_dc_resistance}. The contribution from the bond wires and measurement lines was neglected, since these resistances are small compared with the normal-state resistance of the device.

From the resistance just above the superconducting transition, the normal-state sheet resistance was estimated to be approximately $R_{\square} \simeq 18~\Omega/\square$. This value was used in Fig.~\ref{fig:3}(a) to estimate the kinetic inductance contribution. The transition temperature extracted from this DC measurement, $T_c \simeq 13.4~\mathrm{K}$, is also consistent with the value obtained from the microwave fitting in Fig.~\ref{fig:3}(a).

\section{\label{supp:vortex}Details of vortex-penetration scale}
Thin superconducting strips in perpendicular field exhibit geometry-dependent vortex entry, trapping, and loss, with characteristic field scales that depend strongly on strip width~\cite{likharev1971mixedstate,maksimova1998mixedstate,brandt1993strip,kuit2008vortextrapping,stan2004vortexexpulsion,bulaevskii2011vortexdissipation}. In particular, the zero-field-cooled vortex-penetration scale for a thin strip can be written phenomenologically as
\begin{equation}
\label{eq:Bq}
B_Q^{(j)}
\propto
\frac{B_v + C}{\sin\theta_j},
\qquad
B_v =
\frac{\Phi_0}{2\pi \xi_{\mathrm{eff}} w}.
\end{equation}
Here, $j$ labels the measurement session (since subtle differences in sample preparation and mounting within the cryostat lead to small changes in alignment with the field), $\Phi_0$ is the superconducting flux quantum, $\xi_{\mathrm{eff}}$ is an effective coherence length, $\theta_j$ is the effective misalignment angle between the applied field and the film plane, and $C$ is a phenomenological offset accounting for static perpendicular-field offsets and bond-pad contributions to vortex nucleation (assumed to be roughly constant across designs), lowering $B_Q$. The zero-field-cooled entry scale in Eq. \ref{eq:Bq} should be distinguished from the field-cooled vortex-trapping scale, which scales approximately as $\Phi_0/w^2$~\cite{stan2004vortexexpulsion,kuit2008vortextrapping,kuit2009vortextrapping,roitman2024vortexloss}.

\section{\label{supp:min_dout}Minimum spiral diameter for a target inductance}

\begin{figure}[t]
    \centering
    \includegraphics[width=0.7\columnwidth]{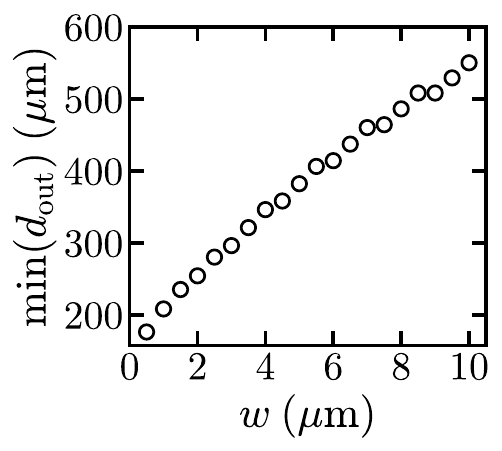}
   \caption{\label{fig:min_dout}
Calculated minimum outer spiral diameter required to obtain $L=100~\mathrm{nH}$ as a function of linewidth $w$, for fixed $d_{\mathrm{in}}=50~\mu\mathrm{m}$ and $L_{\mathrm{k,\square}}=1.74~\mathrm{pH/sq}$.}
\end{figure}

To estimate the minimum spiral footprint required for a target inductance of $100~\mathrm{nH}$, we numerically solved
\begin{align}
    L(n,w,d_{\mathrm{in}}) & =
    L_{\mathrm{g}}(n,w,d_{\mathrm{in}})
    +
    L_{\mathrm{k}}(l,w)
    \\&=
    100~\mathrm{nH}, \notag
\end{align}
  
where $L_{\mathrm{g}}$ is the geometric inductance described by Eq.\ref{eq:Lg_supp} and $L_{\mathrm{k}}=L_{\mathrm{k,\square}}N_{\mathrm{\square}}$ is the kinetic-inductance contribution. Using $L_{\mathrm{k,\square}}=1.74~\mathrm{pH/sq}$ and fixing $d_{\mathrm{in}}=50~\mu\mathrm{m}$ and $g = 2~\mu\text{m}$, the required number of turns was found numerically for each linewidth $w$, and the corresponding minimum outer diameter $d_{\mathrm{out}}$ was calculated. Solving for a given $L$ shows that the selection of a narrower track width allows for better miniaturisation. The plot of minimum calculated $d_{\text{out}}$ as a function of width under these constraints is plotted in Fig.~\ref{fig:min_dout}, demonstrating that reduced width enables reduced footprint in addition to magnetic-field resilience up to higher fields.

\section{\label{supp:BT}Extended data-sets}

\begin{figure*}[t!]
    \centering
    \includegraphics[width=1\textwidth]{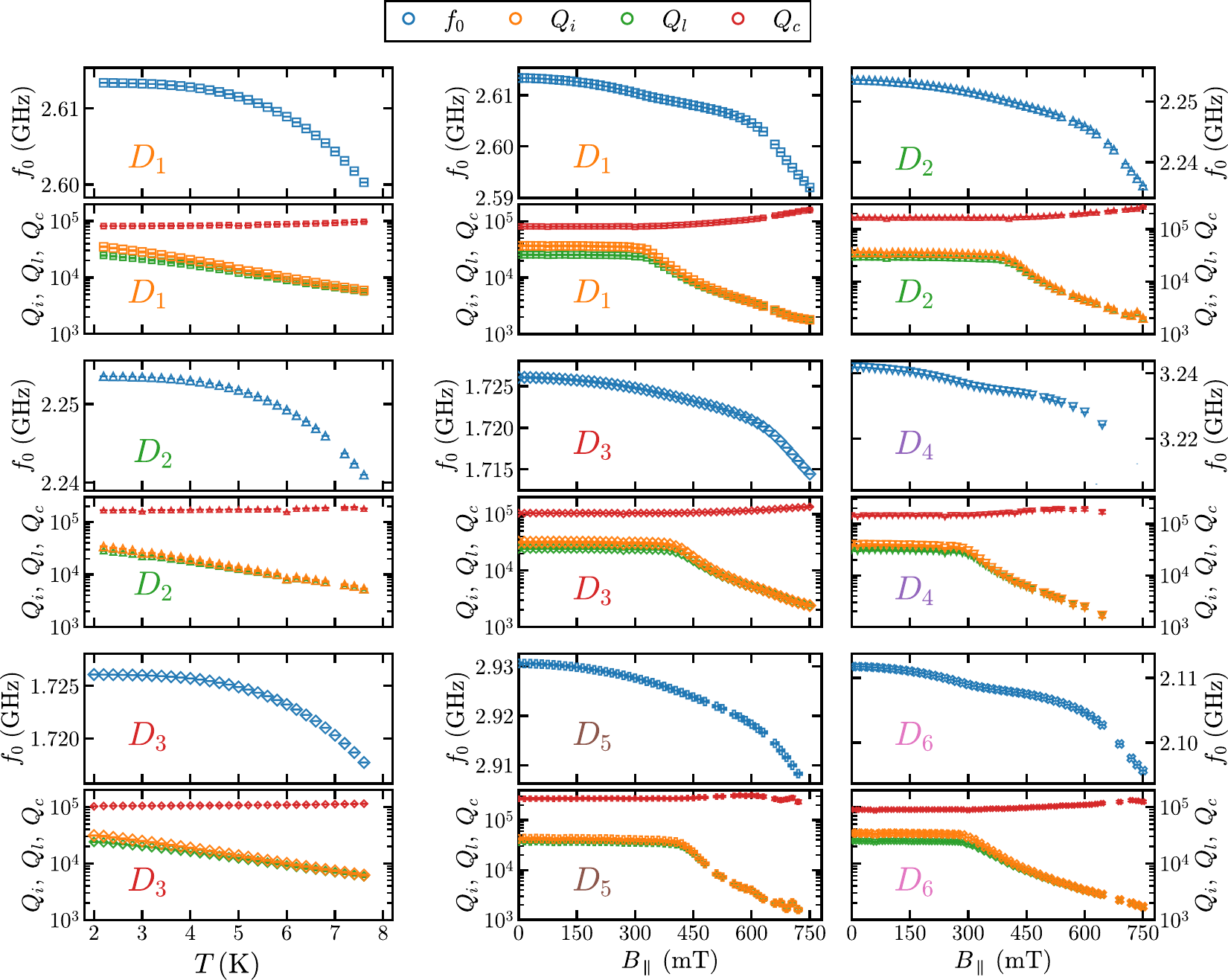}
    \caption{\label{fig:supp_BTdep}
    Temperature and
magnetic-field dependence of $f_0$, $Q_i$, $Q_l$, and $Q_c$ for superconducting spiral resonators $D_{1}$--$D_{3}$ (temperature) and $D_{1}$--$D_{6}$  (magnetic-field).}
\end{figure*}

Fig.~\ref{fig:supp_BTdep} shows the temperature- and magnetic-field-dependent microwave response measured using set-up~1, shown in Fig.~\ref{fig:2}(b--d). Device geometries and design parameters are listed in Table~\ref{tab:spiral_devices}.

\clearpage
\bibliography{ref_list}

\end{document}